\documentclass[fleqn,usenatbib]{mnras}

\usepackage[T1]{fontenc}
\usepackage{ae,aecompl}
\usepackage[normalem]{ulem}

\usepackage{graphicx}	
\usepackage{amsmath}	
\usepackage{amssymb}	
\usepackage{etoolbox}
\makeatletter
\patchcmd\@combinedblfloats{\box\@outputbox}{\unvbox\@outputbox}{}{%
   \errmessage{\noexpand\@combinedblfloats could not be patched}%
}%
 \makeatother
\newcommand{\code}[1]{\texttt{#1}}

\input{vectors}
\usepackage[usenames,dvipsnames]{color}
\usepackage[]{color}

\newcommand{\arx}[1]{arXiv:1809.03666v1 [astro-ph.SR]}
\newcommand{\jphg}[1]{Journal of Physics G}

\title[The impact of (n,$\gamma$) rate uncertainties]{The impact of (n,$\gamma$) reaction rate uncertainties on the predicted abundances of i-process elements with $32\leq Z\leq 48$ in the metal-poor star HD94028}

\author[J. E. McKay et al.]{John E. McKay$^{1,2,\dagger}$,
Pavel A. Denissenkov,$^{1,3,\dagger}$\thanks{E-mail: pavelden@uvic.ca}
Falk Herwig,$^{1,3,\dagger}$, Georgios
\newauthor
Perdikakis$^{3,4,5}$
and Hendrik Schatz$^{3,5,6,\dagger}$
\\
$^{1}$Department of Physics \& Astronomy, University of Victoria, Victoria, B.C., V8W~2Y2, Canada\\
$^{2}$TRIUMF, 4004 Wesbrook Mall, Vancouver, B.C., V6T~2A3, Canada\\
$^{3}$Joint Institute for Nuclear Astrophysics, Center for the Evolution of the Elements, Michigan State University, 640 South Shaw Lane,\\
   East Lansing, MI 48824, USA\\
$^{4}$Department of Physics, Central Michigan University, Mt. Pleasant, Michigan 48859, USA\\
$^{5}$National Superconducting Cyclotron Laboratory, Michigan State University, East Lansing, MI 48824, USA\\
$^{6}$Department of Physics \& Astronomy, Michigan State University, East Lansing, Michigan 48824, USA\\
$^\dagger$NuGrid Collaboration, \href{http://nugridstars.org}{http://nugridstars.org}\\}

\date{Accepted XXX. Received YYY; in original form ZZZ}

\pubyear{2019}

\begin{document}
\label{firstpage}
\pagerange{\pageref{firstpage}--\pageref{lastpage}}
\maketitle

\begin{abstract}
Several anomalous elemental abundance ratios have been observed in the metal-poor star HD94028.
We assume that its high [As/Ge] ratio is a product of a weak intermediate (i) neutron-capture process.
Given that observational errors are usually smaller than predicted nuclear physics uncertainties,
we have first set up a benchmark one-zone i-process nucleosynthesis simulation results of which
provide the best fit to the observed abundances.
We have then performed Monte Carlo simulations in which 113 relevant (n,$\gamma$) reaction rates of unstable species were randomly
varied within Hauser-Feshbach model uncertainty ranges for each reaction to estimate the impact on the predicted stellar abundances.
One of the interesting results of these simulations is a double-peaked distribution of the As
abundance, which is caused by the variation of the $^{75}$Ga (n,$\gamma$) cross section.
This variation strongly anti-correlates with the predicted As abundance,
confirming the necessity for improved theoretical or experimental bounds on this cross section.
The $^{66}$Ni (n,$\gamma$) reaction is found to behave as a major
bottleneck for the i-process nucleosynthesis. Our analysis finds the Pearson product-moment
correlation coefficient $r_\mathrm{P} > 0.2$ for all of the i-process elements with $32 \leq Z \leq 42$, with significant changes
in their predicted abundances showing up when the rate of this reaction is reduced to its theoretically constrained lower bound.
Our results are applicable to any other stellar nucleosynthesis site
with the similar i-process conditions, such as Sakurai's object (V4334 Sagittarii)
or rapidly-accreting white dwarfs.
\end{abstract}




\section{Introduction}
\label{sec:intro}
  
Most of the chemical elements between Fe and Pb have been produced in the slow (s) or/and in the rapid (r) neutron-capture
processes in stars or in stellar explosions. The neutron density in the s process is so low, $N_\mathrm{n}\sim 10^8\ \mathrm{cm}^{-3}$,
that its path nearly follows the n-rich boundary of the valley of stability, outside of which the rates of $\beta$ decay of unstable
isotopes exceed the n-capture rates \citep[e.g.,][]{busso:99,kappeler:11}. 
Because the s-process path is adjacent to the valley of stability,
there are experimental nuclear physics data suitable for its modelling.
On the contrary, the neutron density in the r process
is so high, $N_\mathrm{n}\sim 10^{20}\ \mathrm{cm}^{-3}$ \citep[e.g.,][]{thielemann:11}, that its path approaches the neutron drip line and involves
n-rich unstable isotopes with much less certain n-capture rates from theory. Therefore,
the r-process nucleosynthesis is often taken into account in a chemical composition analysis of
peculiar stars is by using as templates heavy-element abundance patterns observed
in stars that are empirically robustly determined to have been enriched only by an r-process source \citep[e.g.,][]{rpsignature}.

\cite{paper:roederer-2016} tried to reproduce
the anomalous heavy-element abundance distribution in the metal-poor star HD94028 
([Fe/H]\footnote{We use the standard spectroscopic notation [A/B] $ = \log_{10}(N(\mbox{A})/N(\mbox{B}))
- \log_{10}(N_\odot(\mbox{A})/N_\odot(\mbox{B}))$, where $N(\mbox{A})$ and $N_\odot(\mbox{A})$ are abundances of an element A
in a star and the Sun.}$=-1.6$) by a superposition of s- and r-process
enrichments, using the s-process yields from the low-metallicity asymptotic giant branch (AGB) models of \cite{karakas:14}
and \cite{shingles:15} and the r-process abundances from the star HD108317 measured by \cite{roederer:12} and
\cite{roederer:14}, but they did not succeed. In particular, they were not able to explain the high
[As/Ge] and low [Se/As] abundance ratios accompanied by the high [Mo/Fe] and [Ru/Fe] in that star.
After considering possible contributions from other stellar nucleosynthesis sources, such as the weak r-process, neutron- or
proton-rich neutrino winds and $\alpha$-rich freezout in core-collapse supernovae (SNe), electron-capture SNe,
and the s process in fast-rotating massive stars, \cite{paper:roederer-2016} came to the conclusion that
the best fit to the distribution of the heavy-element abundances in HD94028 is obtained assuming that
the s- and r-process abundances are complemented by abundances of the elements with $Z<50$ produced in a weak intermediate (i)
n-capture process, because only the latter seems to be able to provide the observed [As/Ge]\,$\approx +0.99$,
[Se/As]\,$\approx -0.16$ and [Mo/Fe]\,$\approx +0.97$.

The i process in stars was first proposed by \cite{paper:cowan-rose}. It occurs when H is ingested into a He convective zone
at its top by some boundary mixing mechanisms, e.g. by the Kelvin-Helmholtz instability, with convection in the zone driven
by He burning at its bottom. While the ingested H is being transported by convection to the deeper layers with
an increasing temperature, it reacts with the abundant $^{12}$C nuclei via
$^{12}$C(p,$\gamma)^{13}$N at a depth where this reaction becomes as fast as convective mixing.
The radioactive $^{13}$N decays to $^{13}$C, while being carried further down to the bottom of the convective zone,
because its half-life of 9.96 min is comparable to the convective overturn timescale.
Finally, at the bottom, where the temperature is the highest, neutrons are released via $^{13}$C($\alpha$,n)$^{16}$O.
For the H ingestion rates $\dot{M}_\mathrm{ing}\sim 10^{-12}$\,--\,$10^{-9}\ M_\odot\,\mathrm{s}^{-1}$, estimated
for convective He-shell flashes on CO white dwarfs from both 1D stellar evolution models and 3D hydrodynamic
simulations, the i-process neutron density, that is proportional to $\dot{M}_\mathrm{ing}$ but also depends on the metallicity [Fe/H], varies from
$N_\mathrm{n}\sim 10^{13}\ \mathrm{cm}^{-3}$ to $N_\mathrm{n}\sim 10^{16}\ \mathrm{cm}^{-3}$, i.e. it is intermediate
between the values characteristic of the s- and r-processes \citep{paper:herwig-2011,paper:herwig-2014,denissenkov:17,denissenkov:18}.

The i-process action in a real star had most likely been
witnessed by \cite{paper:asplund1999} in the post-AGB star Sakurai's object (V4334 Sagittarii) when it experienced
a very late thermal pulse of its He shell \citep{paper:herwig-2011}.
Possible indirect signatures of the i process are the anomalous heavy-element
abundance distributions in the carbon-enhanced metal-poor (CEMP) r/s stars
\citep{paper:bertoli,paper:dardelet,hampel:16,denissenkov:18}, the Pb deficiency in low-metallicity post-AGB stars \citep{lugaro:15},
the high [Ba/La] ratios in young open clusters \citep{mishenina:15}, and the anomalous isotopic abundances in
some presolar meteoritic grains \citep{jadhav:13,fujiya:13,liu:14}. Stellar evolution theory predicts that,
besides the H-ingesting He-shell flash convection on white dwarfs, the i process may also occur
in metal-poor low-mass thermally-pulsing AGB stars \citep{lugaro:12},
during a He-core flash in metal-poor RGB stars \citep{campbell:10}, in super-AGB stars \citep{jones:16}, and
in Population-III massive stars when their H- and He-burning shells merge \citep{clarkson:19,banerjee:18}.

In this work, we investigate the robustness of the weak i-process elemental abundances and their ratios that are proposed 
to contribute to the anomalous chemical composition of the star HD94028. The predicted abundances are affected by
uncertainties of the (n,$\gamma$) reaction rates for the unstable isotopes involved in this nucleosynthesis. 
By ``weak'' with typical neutron exposure $\tau\sim 1$ we mean an i process that, in spite of having
a high neutron density, does not reach a large neutron exposure (time-integrated neutron flux), e.g. because it
is terminated by the violent, global, non-radial instability that can be induced by H ingestion into 
the He-shell convection \citep{herwig:14}, as probably happened in Sakurai's object
\cite{paper:herwig-2011}. As a result of this ``weakness'' (a low neutron exposure), only the first n-capture peak elements (those with the neutron
number close to the magic one $N=50$) are
produced in a significant amount by the weak i process. We employ the same computational methods and
analysis tools that \cite{paper:dpa-2018} (hereafter, Paper~I) used in a similar study of i process
producing the first peak elements observed in Sakurai's object. 
Unlike in Paper~I, we do not know the i-process site that contributed to the composition of
HD94028. Therefore, we have to use a site-independent one-zone model with the temperature $T$, density $\rho$ and initial chemical
composition adjusted to mimic the i-process physical conditions appropriate for this case,
like it was done by \cite{paper:roederer-2016}. We also deploy constant neutron-density equilibrium models
that are independent of the neutron-source reactions.

\section{Methods}

\subsection{The benchmark simulation}

For the one-zone simulations of the i process nucleosynthesis we use the NuGrid code \code{ppn} \citep{pignatari:16} with the same fixed
temperature $T=2\times 10^8$ K and density $\rho = 10^4\ \mathrm{g\,cm}^{-3}$ as in Paper~I.
The initial chemical composition is prepared using the solar system abundances of \cite{paper:asplund09}
scaled to the metallicity of the star HD94028 ([Fe/H]\,$=-1.6$), assuming that the $\alpha$-element abundances
are enhanced with the mean value of [$\alpha$/Fe] equal to $+0.4$. The initial mass fraction of $^{12}$C
is increased to $X(^{12}\mathrm{C})=0.5$ to be close to the values found in He-flash convective zones.
We also reduce the initial hydrogen abundance to the value of $X(^1\mathrm{H})=0.2$ adjusted to guarantee
that the neutron density will reach the values of $N_\mathrm{n} = 10^{13}$\,--\,$10^{16}\ \mathrm{cm}^{-3}$
typical for the i process. The abundance of $^{16}$O is adjusted accordingly to achieve the required metallicity.
As the comparison with the equilibrium models show, the details of these initial conditions do not impact
the conclusions of this study.

The \code{ppn} simulations are run until the predicted nucleosynthesis yields match the observed abundances as well as possible. 
At this time, the high $N_\mathrm{n}$ phase would have ended because of hydrodynamic feedback from 
the $^{12}$C(p,$\gamma)^{13}$N reaction.
The final abundances obtained in \code{ppn} simulations have to be allowed to decay for
a reasonably long time, say 1 Gyr, after which they may mix with some background abundances
that are not necessarily equal to the initial abundances assumed for the star HD94028 (see Appendix \ref{sec:appendix}).

Our one-zone simulations of i-process nucleosynthesis use the full \code{ppn} network of $\sim 5200$ isotopes
and $\sim 67000$ nuclear reactions. The reaction rates in the default network are taken from several sources
that are all referenced in Paper~I. The NuGrid input physics module provides a possibility to increase or decrease any
included reaction rate by changing its multiplication factor $f_i$ from the
default value $f_i = 1$. This option is used in our Monte Carlo simulations of the impact of
reaction rate uncertainties on the predicted abundances.

The benchmark one-zone simulation uses the above described \code{ppn} code setup with $f_i=1$ and runs until its predicted decayed
elemental abundances match, as well as possible, those observed in HD94028.

\subsection{Reaction Rates \& Maximum Variation Factors}
\label{subsubsec:mc-method-reac}

\begin{figure}
  \centering
  \includegraphics[width=\columnwidth]{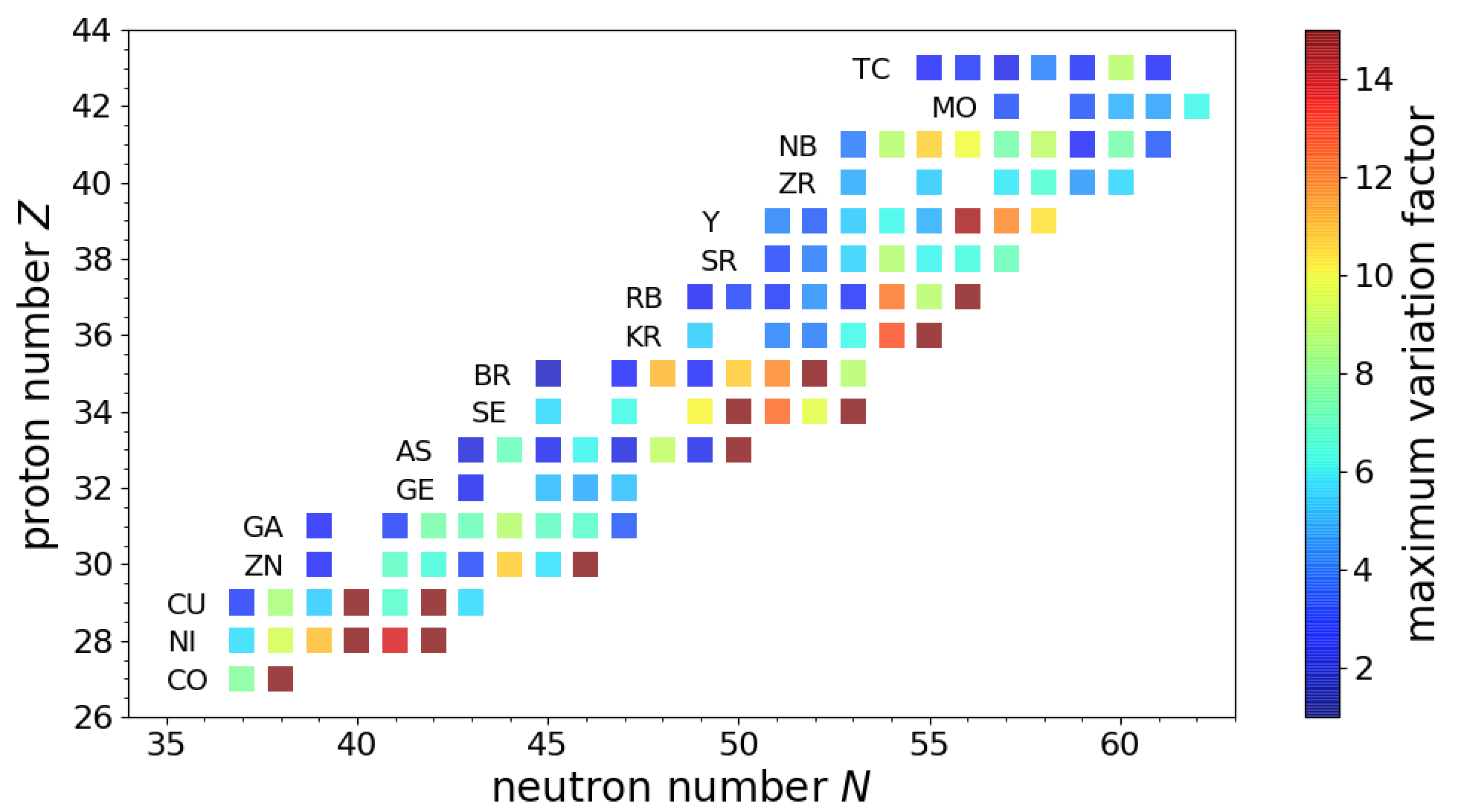}
  \caption{The unstable isotopes whose (n,$\gamma$) reaction rates were varied in this study and the maximum variation
     factors used for the rates. The maximum variation factors for $^{65}$Co, $^{68}$Ni, $^{70}$Ni, $^{69}$Cu, $^{71}$Cu,
     $^{76}$Zn, $^{83}$As, $^{84}$Se, $^{87}$Se, $^{87}$Br, $^{91}$Kr, and $^{93}$Rb exceed the maximum value of
     $v_i^\mathrm{max} = 15$ assigned to the color map.}
  \label{fig:fig0}
\end{figure}

The abundances obtained in the benchmark simulation depend on the (n,$\gamma$) reaction rates
for unstable isotopes. Most of these (n,$\gamma$) rates in the default setup of 
the \code{ppn} code come from the JINA REACLIB v1.1 library \citep{cyburt:10} that recommends
theoretical values calculated using the Hauser-Feshbach model code NON-SMOKER \citep{paper:rauscher}.
However, different Hauser-Feshbach models predict different (n,$\gamma$) rates for a same unstable
isotope (e.g., see Fig. 5 in \citealt{paper:bertoli}), which therefore makes these rates quite uncertain.
To take these uncertainties into account, we follow the same procedure as in Paper~I. 
First, we use charts of n-capture reaction fluxes at the maximum neutron density obtained
in the benchmark simulation to select those unstable isotopes whose
(n,$\gamma$) reaction rate variations can affect the predicted abundances. For each of these isotopes
we find a set of (n,$\gamma$) rates $r_i$ calculated with the Hauser-Feshbach code TALYS\footnote{\url{http://talys.eu}}
\citep{TALYS:07} using 20 different combinations of the nuclear level density and $\gamma$ ray
strength function models listed in Table~1 of Paper~I as input physics data. The rate
uncertainty is assumed to be represented by the ratio of the largest to the lowest rate from this set,
$v_i^\mathrm{max} = r_i^\mathrm{max}/r_i^\mathrm{min}$, that we call the rate's maximum variation factor.
The 113 unstable isotopes selected for the uncertainty study in this
work are displayed with their radiative n-capture rates' maximum varation factors in Figure~\ref{fig:fig0}.

\subsection{The Monte Carlo Simulations}
\label{subsec:mc-method}

Our reaction rate uncertainty study is based on Monte Carlo (MC) simulations in which
we perform 10000 runs of the \code{ppn} code with initial setups that differ from the benchmark simulation only by
different choices of the rate multiplication factors for the selected (n,$\gamma$) reactions.
Each of the MC simulation runs uses a different set of
these factors in which $f_i = (p/v_i^\mathrm{rand}) + (1-p)v_i^\mathrm{rand}$, where $p$ is assigned a value of either 0 or 1
with equal probability, and $v_i^\mathrm{rand}$ is randomly selected from a uniform distribution
between 1 and $v_i^\mathrm{max}$ (Paper~I).

\section{\label{sec:analysis}Results}
  
\subsection{\label{subsec:template}One-zone simulations of the i process}

\begin{figure}
  \centering
  \includegraphics[width=\columnwidth]{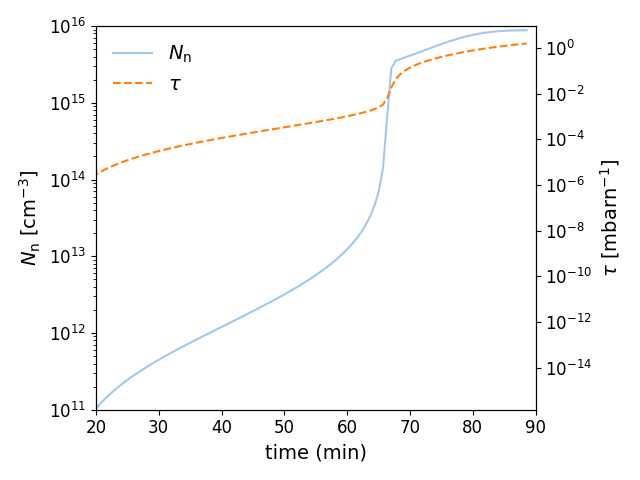}
  \caption{The evolution of the neutron number density and neutron exposure in our benchmark one-zone simulation.}
  \label{fig:fig1}
\end{figure}

Figure \ref{fig:fig1} shows how the neutron number density $N_\mathrm{n}$ changes with time
in our benchmark simulation. It
reaches a value of $\sim 10^{16}\ \mathrm{cm}^{-3}$ at its maximum, indicating
an i-process activation. The final nucleosynthesis yields in the i process also depend on its
duration $t$, or on the neutron exposure
$$
\tau = \int_0^t N_\mathrm{n} v_\mathrm{th} dt,
$$
where $v_\mathrm{th}$ is the thermal velocity of neutrons. The evolution of the neutron exposure
in our benchmark simulation is also shown in Figure \ref{fig:fig1}. In the weak i process the latter
never reaches the values of $\tau\sim 10$\,--\,$100$ at which abundance ratios of neighbouring elements of
the first ($N\approx 50$) and second ($N\approx 82$) n-capture peaks attain their equilibrium values.

\begin{figure}
        \centering
        \includegraphics[width=\columnwidth]{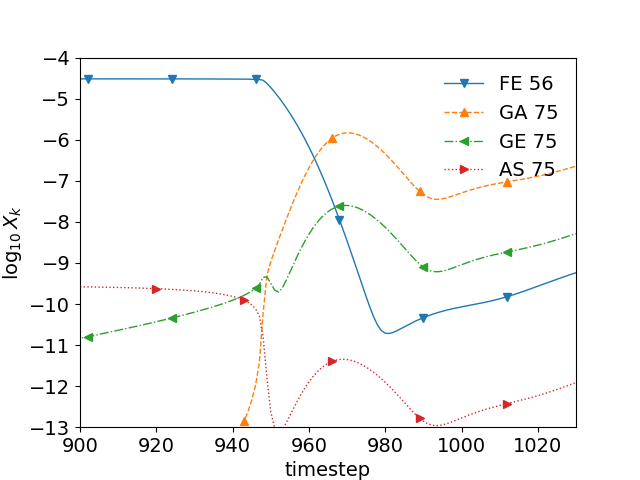}
        \caption{The evolution of the abundances of the two isotopes that decay into As and of the Fe seed
          in the benchmark simulation. The timestep 960 corresponds to 75 minutes of integration time.}
        \label{fig:fig2}
\end{figure}

In order to identify a narrow interval of integration timesteps
or intervals of $t$ and $\tau$ in which the anomalous elemental abundance
ratios observed in the star HD94028 are produced in the weak i process, we first examine the evolution of
the isotopic progenitors of one of our main target elements As in Figure \ref{fig:fig2}. We see that the abundances of
the unstable isotopes $^{75}$Ga and $^{75}$Ge reach their peak values almost immediately after the 960th timestep (in 75 minutes).
If we stop the benchmark simulation here and allow these isotopes to decay into the only stable As isotope $^{75}$As,
we will get its highest possible abundance, which we need to explain the anomalously high
[As/Ge] ratio in the star HD94028.

Figure \ref{fig:fig2} also shows that after the $^{75}$Ga and $^{75}$Ge abundances have reached their peak values
the most abundant of the Fe isotopes $^{56}$Fe ceases being the main seed nucleus for the simulated i process. This is
an artefact of the one-zone model.  Moreover, the very presence of the peaks in the evolutionary profiles of
the $^{75}$Ga and $^{75}$Ge abundances is seen to be caused by the depletion of the Fe abundance.
In multi-zone simulations of the i process in a He convective zone the Fe abundance would not be depleted to such low values,
because Fe destroyed at its bottom by n captures would be replenished by Fe brought there by mixing from
the other parts of the convective zone, and a build-up of $^{75}$Ga and $^{75}$Ge could continue to overall larger values.
However, at the moment when the neutron flux passes through these species they adopt their local equilibrium values.
These are what we are seeking for to compare with the observed abundances.

\begin{figure}
        \centering
        \includegraphics[width=\columnwidth]{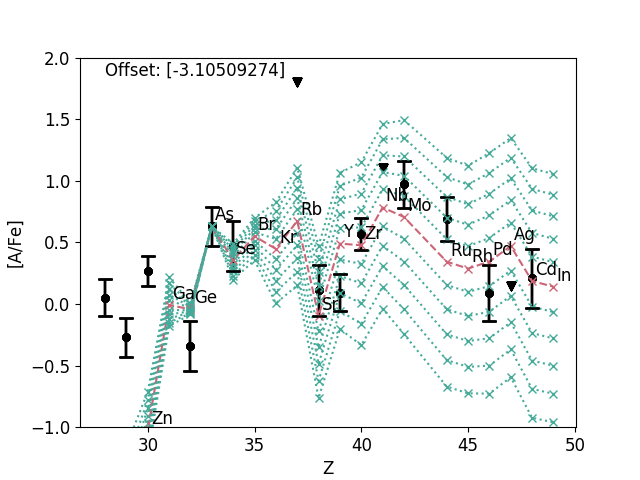}
        \caption{The distributions of $[\mathrm{A}/\mathrm{Fe}]_\mathrm{theor}$ predicted by the benchmark simulation for the timesteps
    from 968 through to 978 (from the lower to upper curve) are compared with the observed values of
    $[\mathrm{A}/\mathrm{Fe}]$ from \protect\cite{paper:roederer-2016} to determine
    the best-fit timestep for the star HD94028. The 973rd timestep on the 85th minute has the minimum value of $\chi^2 = 18.0$ (the brown-dashed curve).
    The values of $[\mathrm{As}/\mathrm{Fe}]_\mathrm{theor}$ are all pinned to the observed [As/Fe] ratio (see Appendix A).
    The upside-down triangles are the upper limits.}
        \label{fig:fig3}
\end{figure}

Next, we use Figure \ref{fig:fig3} to determine a timestep at which the decayed and mixed elemental abundances predicted by
the benchmark model best fit their corresponding i-process abundances derived by \cite{paper:roederer-2016}
for the star HD94028. With the predicted abundance of As pinned to its observed value
the best-fit timestep turns out to be the 973rd one that corresponds to 85 minutes of integration time.
At this time the calculated abundances $[\mathrm{A}/\mathrm{Fe}]_\mathrm{theor}$  minimize
$$
\chi^2 = \sum_{32\leq Z\leq 48}\frac{\left([\mathrm{A}/\mathrm{Fe}]_\mathrm{obs} - [\mathrm{A}/\mathrm{Fe}]_\mathrm{theor}\right)^2}{
\sigma\left([\mathrm{A}/\mathrm{Fe}]_\mathrm{obs}\right)^2},
$$
where $Z$ is the proton number of an element A with available observed abundance $[\mathrm{A}/\mathrm{Fe}]_\mathrm{obs}$  
and corresponding error $\sigma\left([\mathrm{A}/\mathrm{Fe}]_\mathrm{obs}\right)$.

The normalization factor $d\approx -3.1$dex (Offset in Figure~\ref{fig:fig3}) used to match the abundance pattern at As gives 
an indication of the fraction of i-process isotopes in the composition (see Appendix A). 

\begin{figure}
        \centering
        \includegraphics[width=\columnwidth]{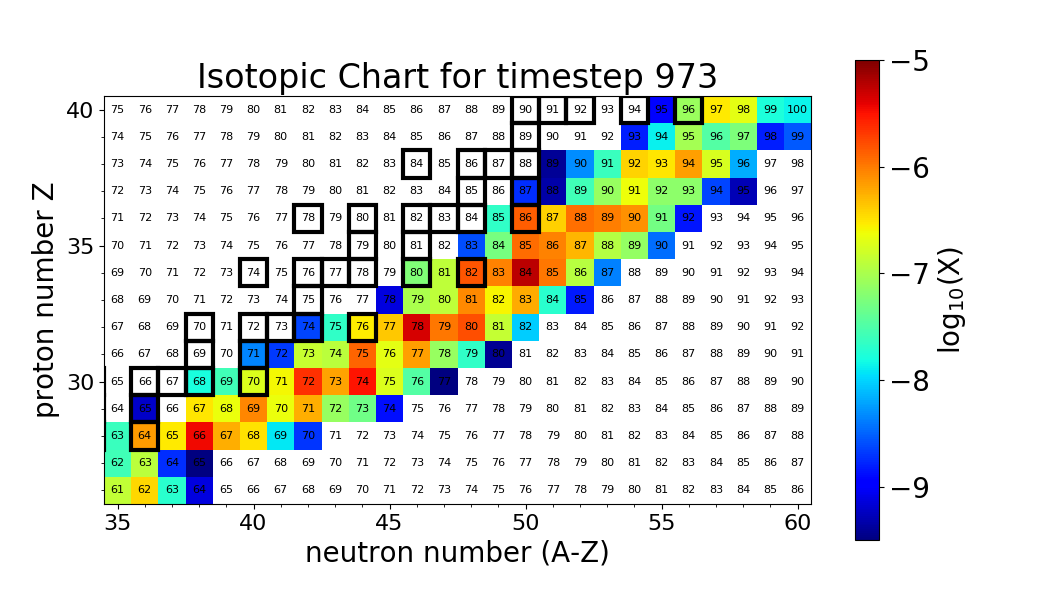}
        \caption{Undecayed isotopic abundances (mass fractions $X$) for the best-fit timestep 973 near the i-process nucleosynthesis peak.}
        \label{fig:fig4}
\end{figure}

A characteristic feature of an i-process is that its path band on a chart of nuclides involves only unstable isotopes that are two to eight
neutron numbers away from the valley of stability (Figure~\ref{fig:fig4}). During the i process,
especially large abundances are usually found for isotopes with neutron numbers close to the magic ones.
In our case these are $^{84}$Se, $^{85}$Br and $^{86}$Kr with $N=50$ and $34\leq Z\leq 36$.
Large abundances also occur for the unstable Ge isotopes ($Z=32$), $^{78}$Ge, $^{79}$Ge, \& $^{80}$Ge.

Of particular importance for this work are the large abundances of $^{74}$Zn and $^{75}$Ga ($Z=30$ and $31$),
which are significantly higher than the subsequent neutron capture products $^{75}$Zn and $^{76}$Ga.
The latter isotope is the main source of As. This indicated a significant
bottleneck for the n-capture reaction flux at this point.
The Monte Carlo simulations confirm this (see Section \ref{subsec:mc_analysis}).

A notable outlier is $^{66}$Ni ($Z=28$) that has a very high abundance compared to its surrounding isotopes.
Its (n,$\gamma$) reaction product $^{67}$Ni has a lower abundance which, combined with the two-day half-life of $^{66}$Ni,
signifies that $^{66}$Ni also represents an important i-process bottleneck deserving a special attention.

\subsection{The Monte Carlo simulations}
\label{subsec:mc_analysis}

The Monte Carlo (MC) simulations produce 10000 sets of the elemental and isotopic abundances
from the timestep 973 and subsequent 1 Gyr beta decay that provides the best-fit of the theoretical to observed abundances for the benchmark model
(Figure \ref{fig:fig3}). Mixing with other material does not need to be considered here
because we are only interested in finding correlations between the randomly varied reaction rate multiplication factors $f_i$ and
the predicted abundances (mass fractions) $X_k$. These correlations are evaluated using the Pearson product-moment
correlation coefficient $r_\mathrm{P}(f_i,X_k)$ (Equation (2) in Paper~I).

\begin{figure}
        \centering
        \includegraphics[width=\columnwidth]{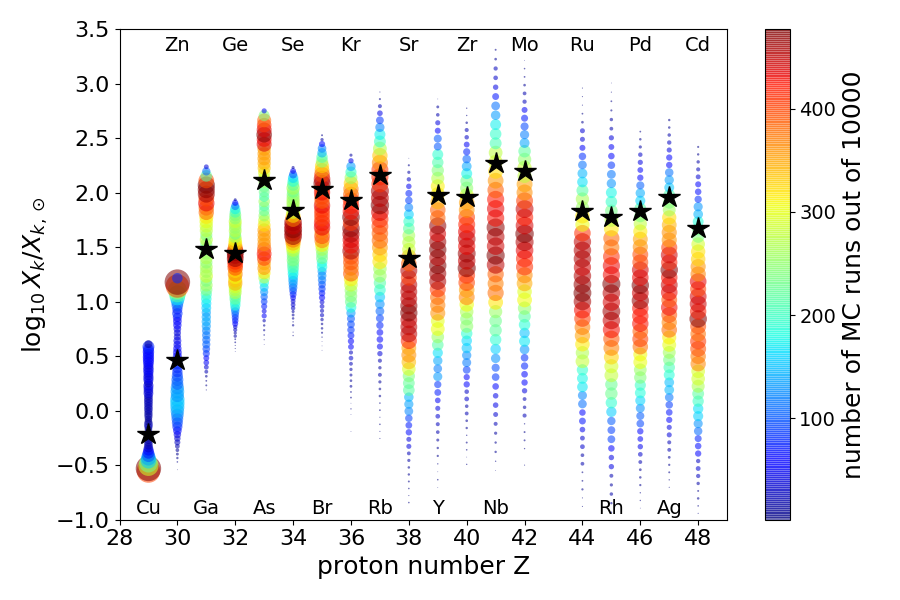}
        \caption{
     The distributions of the i-process elemental abundances
     generated in the Monte Carlo simulation by randomly varying (n,$\gamma$) reaction rates of the unstable isotopes
     that are displayed in Figure~\ref{fig:fig0}.
     The larger and redder circles correspond to a larger number of MC runs contributing to a given abundance.
     The black star symbols show the benchmark simulation abundances.}
        \label{fig:fig6}
\end{figure}

\begin{figure}
        \centering
        \includegraphics[width=0.8\columnwidth]{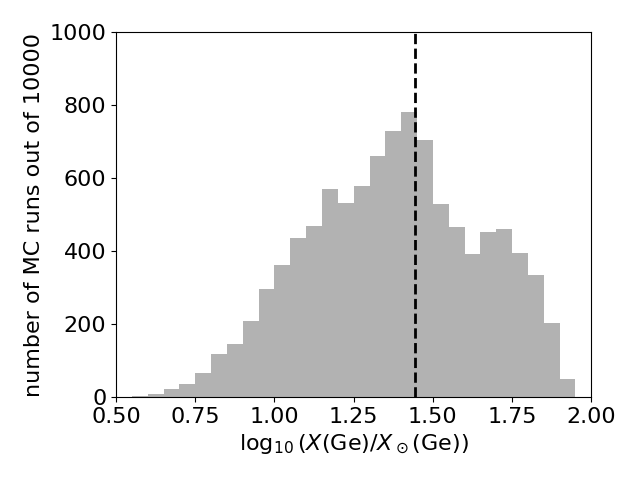}
        \includegraphics[width=0.8\columnwidth]{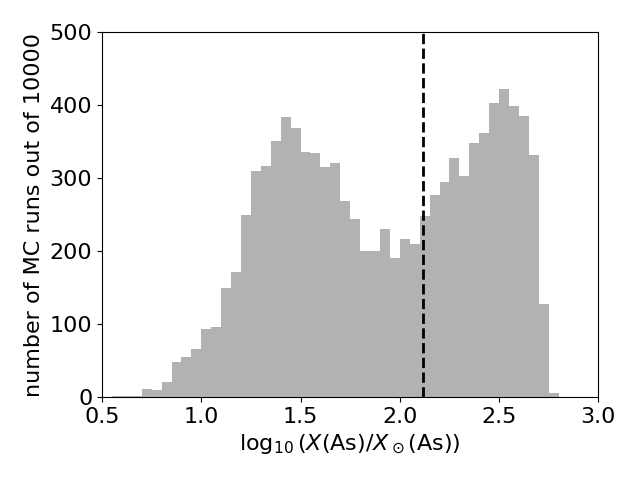}
        \caption{The Ge and As elemental abundance distributions from our Monte Carlo simulation. The abundances from the benchmark
     simulation are marked by black vertical dashed lines.}
        \label{fig:fig7}
\end{figure}

Figure \ref{fig:fig6} shows the final distributions of $\log_{10}(X_k/X_{k,\,\odot})$ obtained in our MC simulation.
Strong deviations from a Gaussian indicate the presence of
important branching points on the i-process paths leading to the synthesis of isobars that decay into the same element.
A prominent example of such a deviation is As whose MC abundances pile up in two distinct groups, one having the high mean
abundance at a level of $\sim 1$dex above the Ge abundances, as indicated by observations, and the other lying roughly on the same level with Ge.
This is shown more clearly in Figure~\ref{fig:fig7}, together with the MC abundance distribution for Ge, 
which also shows a double peaked structure, albeit less pronounced. As we will show below,
the bifurcation of the As abundance distribution is caused by the $^{75}$Ga bottleneck.

\begin{figure}
        \centering
        \includegraphics[width=\columnwidth]{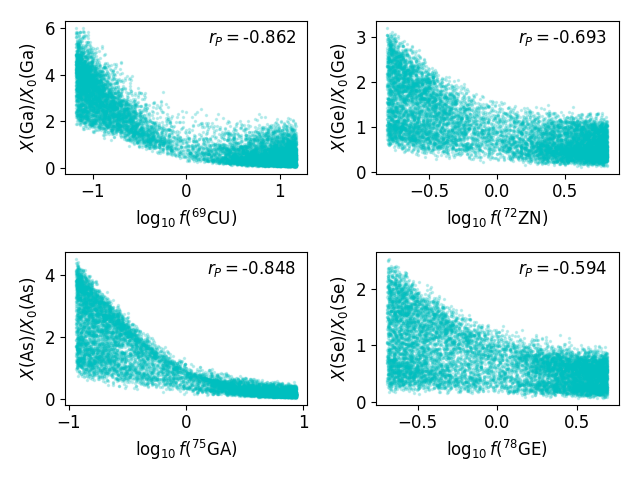}
        \caption{The distributions of the elemental abundances of Ga, Ge, As and Se as functions of the (n,$\gamma$) reaction
    rate multiplication factors for the selected isotopes of Cu, Zn, Ga and Ge for which the absolute magnitudes of
    the correlation coefficient are the largest (shown in the top-right corners of the panels).}
        \label{fig:fig8}
\end{figure}

Figure \ref{fig:fig8} shows the distributions of the Ga, Ge, As and Se abundances as functions of the multiplication
factors for the $^{69}$Cu, $^{72}$Zn, $^{75}$Ga and $^{78}$Ge n-capture rates with which these abundances have the
strongest correlations.

\begin{figure}
        \centering
        \includegraphics[width=\columnwidth]{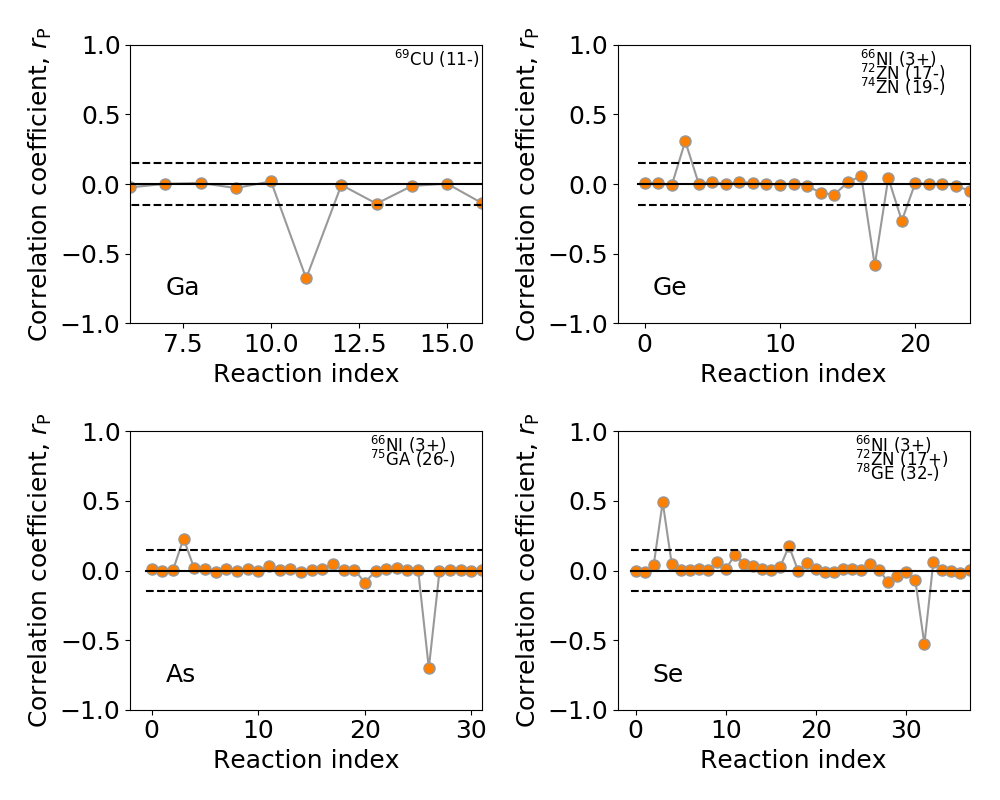}
        \caption{Correlation coefficients for the abundances of Ga (upper left), Ge (upper right), As (lower left), 
                 and Se (lower right)  with reaction rate variations as functions of reaction index. 
                 Reaction indices with a sign of correlation and corresponding neutron capture target isotopes are 
                 given in the legends for the largest correlations.}
        \label{fig:fig9}
\end{figure}

To identify all neutron capture rate uncertainties that affect the relevant elemental abundances,
we have calculated the Pearson correlation coefficient for each of
these elements and for all the isotopes whose (n,$\gamma$) reaction rates were varied.
Figure \ref{fig:fig9} shows the resulting correlations of the predicted Ga, Ge, As and Se
abundances with the (n,$\gamma$) rate multiplication factor for $^{66}$Ni and some other unstable isotopes.
When examining all the i-process elements up to Mo ($Z=42$), we find that $f(^{66}\mathrm{Ni})$
positively correlates with the abundances of every element heavier than Ga (Table~\ref{table:tab3}).
This confirms that $^{66}$Ni is also a major bottleneck isotope.

\begin{table*}
        \centering
        \begin{tabular}{|c|c|c|c|c||c|}
                \hline
                Reaction & Element & $r_\mathrm{P}$ (1-zone, 973rd timestep) & $r_\mathrm{P}$ (1-zone, 979th timestep) & 
                        $r_\mathrm{P}\ (N_\mathrm{n}=10^{16}\,\mathrm{cm}^{-3})$ &
                        $r_\mathrm{P}\ (N_\mathrm{n}=10^{15}\,\mathrm{cm}^{-3})$\\
                \hline\hline
                {$^{66}$Ni}
                & Zn & -0.7793 &  -0.7108  & -0.7497 & -0.7948\\
                & Ge & 0.3079 & -0.1255 & 0.1384 & 0.2286\\
                & As & 0.2298 & -0.0583 & 0.1387 & 0.1969\\
                & Se & 0.4922 & 0.1412 & 0.4309 & 0.5210\\
                & Br & 0.4391 & 0.1340 & 0.3862 & 0.4240\\
                & Kr & 0.5031 & 0.3807 & 0.4938 & 0.6293\\
                & Rb & 0.4130 & 0.3387 & 0.3984 & 0.5215\\
                & Sr & 0.3601 & 0.3133 & 0.3475 & 0.4463\\
                & Y  & 0.3093 & 0.2826 & 0.2929 & 0.4427\\
                & Zr & 0.4021 & 0.4435 & 0.3682 & 0.4646\\
                & Nb & 0.2906 & 0.2905 & 0.2706 & 0.3490\\
                & Mo & 0.3583 & 0.4046 & 0.3174 & 0.3919\\
                \hline
                $^{69}$Cu & Ga & -0.6776 & -0.6071 & -0.6500 & -0.6022\\
                \hline
                $^{72}$Zn & Ge & -0.5842 & -0.6450 & -0.5892 & -0.5943\\
                \hline
                $^{75}$Ga & As & -0.7021 & -0.7291 & -0.7040 & -0.7725\\
                \hline
                $^{78}$Ge & Se & -0.5292 & -0.7188 & -0.5636 & -0.5308\\
                \hline
        \end{tabular}
        \caption{The strongest correlations between the (n,$\gamma$) reaction rate variations and
           the i-process elemental abundances found in our MC simulations.}
        \label{table:tab3}
\end{table*}

\begin{figure}
        \centering
        \includegraphics[width=\columnwidth]{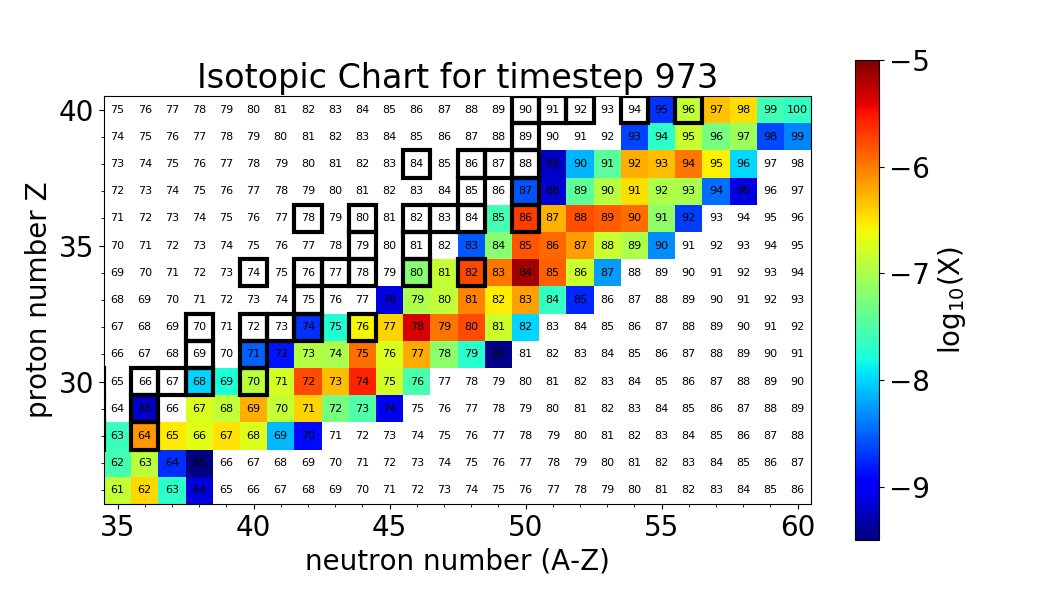}
        \includegraphics[width=\columnwidth]{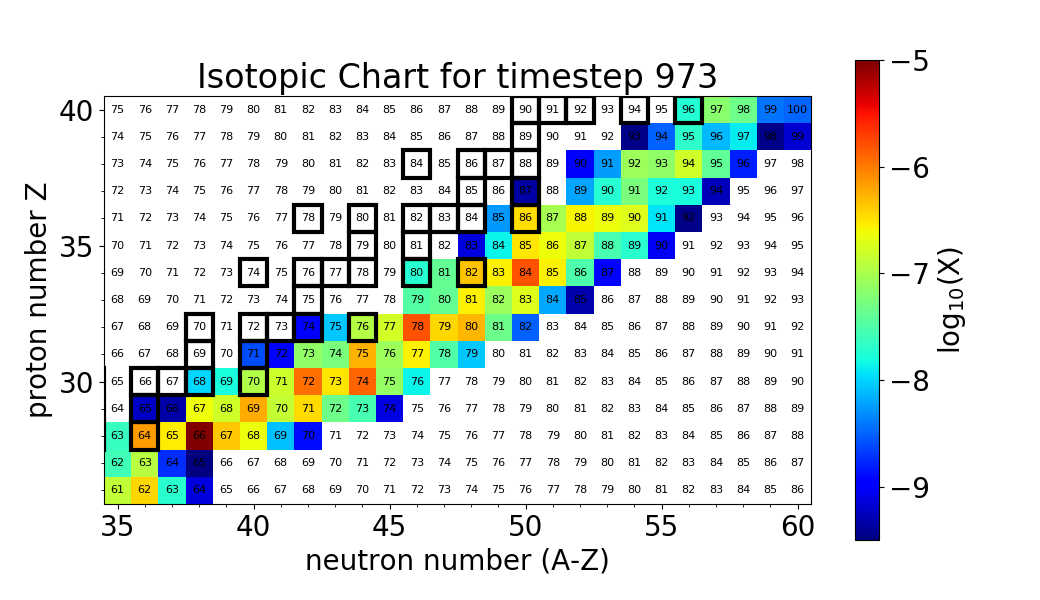}
        \caption{Isotopic abundances from the two additional \code{ppn} runs in which only the multiplication factor
     $f(^{66}\mathrm{Ni})$ was switched between its maximum (top panel) and minimum (bottom panel) values constrained by the Hauser-Feshbach
     model computations.}
        \label{fig:fig10}
\end{figure}

To investigate the impact of the $^{66}$Ni neutron capture rate on the i-process reaction path band, 
we performed two additional calculations, where we only varied $f(^{66}\mathrm{Ni})$ setting it 
to its maximum ($9.4$) and minimum ($0.11$) value, respectively. 
The results are shown in Figure \ref{fig:fig10}.
The maximum $f(^{66}\mathrm{Ni})$ case shows very high isotopic abundances far along the i-process path band,
while in the minimum case the (n,$\gamma$) reaction flux appears to be stuck at $^{66}$Ni,
resulting in a much enhanced accumulation of its abundance.
Many isotopic abundances that had values of
$\log_{10}X_k\geq -6$ in the benchmark simulation have dropped by $\sim 1$--$2$ orders of magnitude.
This result is consistent with
our correlation analysis and emphasizes the role of $^{66}$Ni as a major bottleneck isotope.
Replacing the default value of $f(^{66}\mathrm{Ni})=1$ by $f(^{66}\mathrm{Ni})_\mathrm{max}$ and
$f(^{66}\mathrm{Ni})_\mathrm{min}$ requires shifts of the best-fit timestep from the 973rd to the 972nd
and 976th, respectively. These changes have an almost unnoticeable effect on Figure~\ref{fig:fig10},
therefore they do not affect our conclusion about $^{66}$Ni(n,$\gamma$) being the major bottleneck reaction.

To show that it is the bifurcation of the (n,$\gamma$) reaction flux at the $^{75}$Ga isotope
and not at $^{66}$Ni that
is responsible for the double-peaked distribution of the As abundance, we have divided our MC simulation
runs into two groups, one with $f(^{75}\mathrm{Ga})>1$ and the other with $f(^{75}\mathrm{Ga})<1$.
 \begin{figure}
        \centering
        \includegraphics[width=0.8\columnwidth]{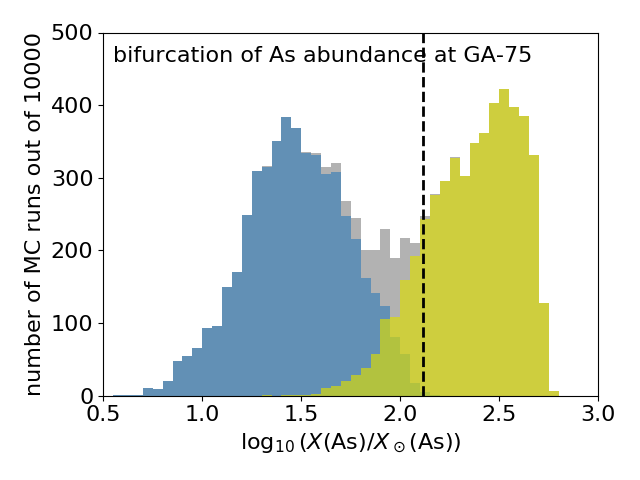}
        \includegraphics[width=0.8\columnwidth]{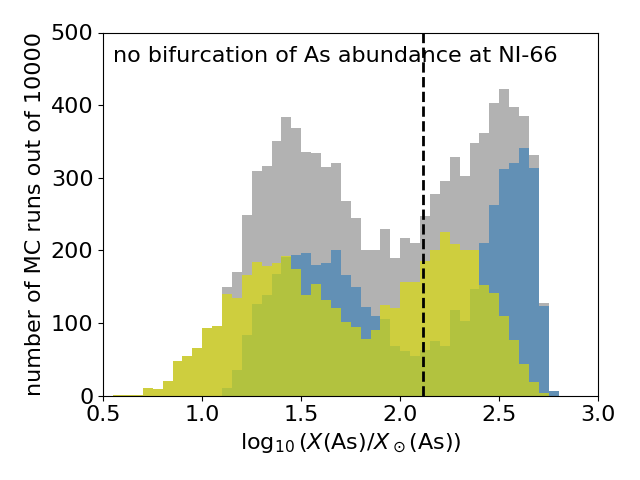}
        \caption{Top panel: the double-peaked distribution of the As elemental abundance in our MC simulation (grey) is decomposed into two
     isolated peaks (blue and yellow) when we divide the As abundances into two groups with $f(^{75}\mathrm{Ga})>1$ and
     $f(^{75}\mathrm{Ga})<1$. Bottom panel: a similar test with $f(^{66}\mathrm{Ni})$ only shifts the distribution to the higher As
     abundances.}
        \label{fig:fig11}
\end{figure}
Indeed this resulted in a separation of the As abundance distribution into two distinct peaks (top panel in Figure \ref{fig:fig11}).
On the other hand, a similar test for $^{66}$Ni neutron capture rate only resulted in a shift of the double-peaked As
abundance distribution to the higher abundance values (bottom panel in Figure \ref{fig:fig11}).
These tests show that although $^{66}$Ni(n,$\gamma$) is the most important reaction for regulating the i-process
nucleosynthesis paths in the $A=75$ region of the chart of nuclides,
the anomalously high abundance ratio [As/Ge] in the star HD94028 is much more strongly affected by the uncertainty of
the $^{75}$Ga(n,$\gamma$) reaction rate that has to be reduced to increase [As/Ge].

The major bottlenecks $^{66}$Ni, $^{72}$Zn, and $^{78}$Ge lie on a sequence of nuclei 
with $Z = 26+n$ and $N = 36+2n$ that for even $n$ have particularly low $Q$-values 
for $\beta^-$ decay ($<1$ MeV) and at the same time relatively low $Q$-values 
for neutron capture ($\sim 5$\,--\,$6$ MeV). This results in both, long half-lives and 
slow neutron capture rates that make these isotopes bottlenecks. 
As the i process begins mostly with the relatively abundant Fe isotopes $^{66}$Ni is 
the first major bottleneck. For the odd $n$ isotopes in this sequence the effect is 
less pronounced but still present --- these are the isotopes $^{69}$Cu, $^{75}$Ga 
for which we also find some sensitivity. 

Figure~\ref{fig:fig6} shows that the neutron capture rate uncertainties have a tendency to reduce abundances below the benchmark values. 
Our analysis reveals that this is
a cumulative effect in which the reaction $^{66}$Ni(n,$\gamma$) plays an important but not defining role.
When using the central values of the distributions for the comparison with observations, 
a fit with similar quality can still be obtained
by increasing the integration time from 85 minutes (the 973rd timestep) to 89 minutes (the 979th timestep). Therefore, we have performed
another MC simulation with the 979th timestep used as the final one and analyzed their results. This analysis
has shown that all our main conclusions based on the results of the MC simulation with the 973rd final (best-fit) timestep remain true.
For example, the values of the last four correlation coefficients
in Table~\ref{table:tab3} have changed less than by 10\% for the first three of them and by 36\% for the fourth one compared to
their values obtained in the first MC simulation (the last four raws in the 3rd and 4th columns). 

We have not adjusted the integration time for each individual run of our MC simulation to get the best fit in order 
to estimate the ``true'' nuclear error bar because what
we understand as nuclear uncertainty contribution to the abundances is 
the variation of abundances in respect to nuclear uncertainties for {\em fixed} astrophysical conditions.

For experimentalists it makes sense because they want to measure anything that affects abundances or 
the choice of stellar physics parameters. For example, if for a change of a bottleneck reaction one needs 
much longer exposure times to get the same abundances this should be flagged as something 
that needs a measurement even if it would result in the same abundance and would not pop up as important in the approach
when the best-fit integration time is adjusted for individual MC runs. 

For comparison with observations we want to include uncertainties of the astrophysical model and yes, 
there can then be correlations that we may miss in our approach, but we think that our simple definition of 
the nuclear component of the error is still sensible even if correlations are present. 
More importantly, we do not really have an astrophysical model for the case of HD94028, but a parametrized approach that is fitted, 
so there are no predictions of astrophysical parameters with errors. This is why it is tempting to use the individual integration time approach
that would give us the nuclear uncertainties that need to be addressed 
to determine whether our parametrized approach can fit the observational data or not 
(which would not be uninteresting --- though there are other parameters to adjust as well). 
Our approach, on the other hand, determines more generally the nuclear errors to expect in an i-process model 
that {\em fixes/predicts} astrophysical parameters (for example any future realistic stellar models) --- 
we just use the parametrized approach to estimate the conditions in such a model ---  
and it also flags reactions that affect the parameter choice. 

The only concern one may have is that the sensitivities we determine depend sensitively 
on the astrophysical parameters, such as the exposure time. This is addressed by doing a second MC analysis 
at a different exposure time that can still fit the data within the nuclear uncertainty.

\subsection{Simulations with a constant neutron density}

To test the robustness of our predictions about the key role of the $^{75}$Ga (n,$\gamma$) cross section in determining the production level of
As and the role of the $^{66}$Ni n-capture reaction as the major bottleneck for the synthesis of most of the i-process elements
in the star HD94028, we have complemented our study with benchmark and MC one-zone simulations in which the neutron number
density was kept constant and hydrogen burning was suppressed by assuming that its mass fraction $X=0$,
like it was done in the work of \cite{hampel:16}.

\begin{figure}
        \centering
        \includegraphics[width=\columnwidth]{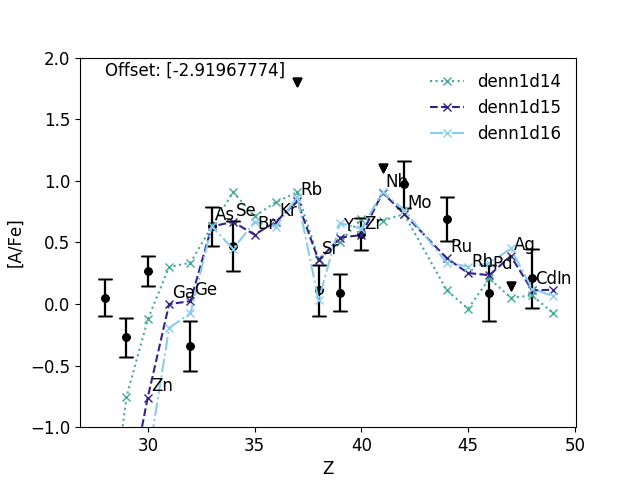}
        \caption{Same as the brown-dashed curve in Figure \ref{fig:fig3}, but for the benchmark one-zone
                 simulations with the constant neutron densities. The best-fit timesteps are
                 1183 (the 688th minute), 985 (the 96th minute) and 787 (the 13th minute)
                 and their corresponding $\chi^2$ values are 38.7, 19.8 and 22.5
                 from the lowest to the highest indicated values of $N_\mathrm{n}$.}
        \label{fig:3denn}
\end{figure}

The results of these simulations are presented in Figure \ref{fig:3denn} and in the last two columns of Table \ref{table:tab3}.
The figure shows that the values of $N_\mathrm{n} < 10^{15}\ \mathrm{cm}^{-3}$ should not be considered because they result
in too low [As/Ge] and too high [Se/As] abundance ratios accompanied by a too low [Ru/Fe] abundance. As for the values of
$N_\mathrm{n} = 10^{15}\ \mathrm{cm}^{-3}$ and $N_\mathrm{n} = 10^{16}\ \mathrm{cm}^{-3}$, the analyses of the results of
MC simulations performed for them have led us to the same conclusions that we made using the one-zone model with
the neutron density evolution profile shown in Figure \ref{fig:fig1}, even the Pearson correlation coefficients have not changed
much (Table \ref{table:tab3}).

\section{Summary and Conclusions}
\label{sec:conclusions}
  
Our study has been focused on the anomalously high [As/Ge] abundance ratio in the
metal-poor star HD94028. Following \cite{paper:roederer-2016}, we have assumed
that this and the other abundance anomalies for the elements with $32 < Z < 48$ were
contributed to this star by a weak i process that had occurred at an unknown stellar site.

\begin{figure}
        \centering
        \includegraphics[width=0.8\columnwidth]{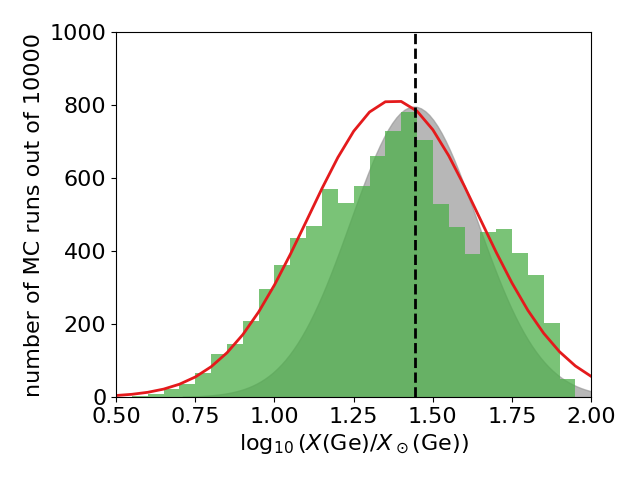}
        \includegraphics[width=0.8\columnwidth]{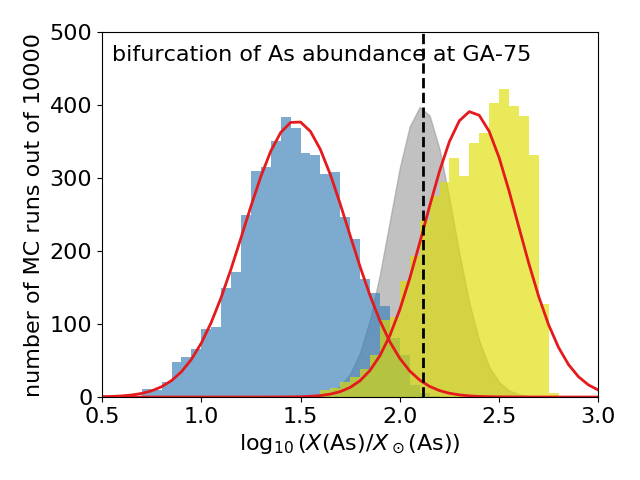}
        \caption{Comparison of the Gaussians fittted to the Ge and bifurcated As abundance distributions in the MC simulation (red curves) 
     with the observed distributions of Ge and As in HD94028 also represented by the Gaussians (grey) with the mean values equal to the Ge and As
     abundances in the benchmark simulation and $\sigma(\mathrm{[Ge/Fe]}_\mathrm{obs})$ and $\sigma(\mathrm{[As/Fe]}_\mathrm{obs})$
     estimated by \protect\cite{paper:roederer-2016}.}
        \label{fig:fig12}
\end{figure}

To identify the isotopes whose (n,$\gamma$) reaction rate uncertainties have the strongest
impact on the predicted As abundance, we have performed Monte Carlo (MC) simulations in which
the (n,$\gamma$) rates of 113 unstable isotopes that could affect
the predicted i-process abundances in the star HD94028 were randomly varied
within their minimum and maximum limits constrained by the Hauser-Feshbach model computations.

The MC simulations are based on the one-zone benchmark model of the i process nucleosynthesis
for which we have selected the integration times (the final timesteps) providing the best fit
of the predicted decayed elemental abundances to
the i-process abundances in HD94028 derived by \cite{paper:roederer-2016}.

The analysis of the results of our MC simulations has revealed that the $^{75}$Ga(n,$\gamma$)
reaction has the strongest impact on the predicted As abundance. 
We have shown that the uncertainty of its rate leads to the bifurcation of the As abundance
distribution. In the bottom panel of Figure~\ref{fig:fig12} we have added
Gaussians (red curves) fitting the two distinct As abundance distributions that are revealed when
we divide the results of MC simulations into the groups with $f(^{75}\mathrm{Ga})<1$ and $f(^{75}\mathrm{Ga})>1$.
We have also added a Gaussian (grey) with a mean value equal to the As abundance from the benchmark
simulation (the black-dashed line) and the standard deviation $\sigma(\mathrm{[As/Fe]}_\mathrm{obs})$ 
estimated for HD94028 by \cite{paper:roederer-2016}. The top panel shows a similar plot, but for Ge.
This figure confirms the conclusion made by \cite{paper:dpa-2018} that nuclear uncertainties of
(n,$\gamma$) reaction rates relevant for i process are usually
significant and overall similar to observational uncertainties. Given that $r_\mathrm{P}(f(^{75}\mathrm{Ga}),X(\mathrm{As}))$
is negative (Table~\ref{table:tab3}), the bottom panel of Figure~\ref{fig:fig12} also shows that
a reduction of the rate of the $^{75}$Ga(n,$\gamma$) reaction
would lead to the desired increase of [As/Ge], while a significant increase of its rate
would rule out the weak i process as one contributing to the abundance anomalies in the star HD94028.

It has also been shown that $^{66}$Ni is the major bottleneck isotope that strongly affects most of the predicted
abundances in the simulated i process. Therefore, experimental measurements of
$^{75}$Ga and $^{66}$Ni n-capture rates would significantly improve our understanding of
the i-process nucleosynthesis and its possible contribution to the elemental abundance anomalies
in the star HD94028 and in other similar objects.

\section*{Acknowledgements}

FH acknowledges funding from NSERC through a Discovery Grant. This
research is supported by the National Science Foundation (USA) under
Grant No. PHY-1430152 (JINA Center for the Evolution of the Elements).
The authors thank Iris Dillmann, Barry Davids and Chris Ruiz for fruitfull
discussions of this problem.
This research was enabled in part by support provided by Compute Canada.
We appreciate the work of many researchers who are involved in the development of
NuGrid computer codes that have been used in this study.




\bibliographystyle{mnras}
\bibliography{paper.bib}



\appendix

\section{Comparison of predicted and observed abundances}
\label{sec:appendix}
  
For the comparison in Figure \ref{fig:fig3} we have used the standard spectroscopic ratios of the observed
abundances in the star HD94028
$$
[\mathrm{A}/\mathrm{Fe}]_\mathrm{obs} = \log_{10}\frac{X_\mathrm{A,obs}}{X_\mathrm{A,\odot}} -
\log_{10}\frac{X_\mathrm{Fe,obs}}{X_{\mathrm{Fe},\odot}},
$$
where $X_\mathrm{A}$ is the mass fraction of an element A.
Their theoretical counterparts are
$$
[\mathrm{A}/\mathrm{Fe}]_\mathrm{theor} = \log_{10}\frac{X_\mathrm{A,theor}}{X_{\mathrm{A},\odot}}-
\log_{10}\frac{X_\mathrm{Fe,theor}}{X_{\mathrm{Fe},\odot}},
$$
where
$$
X_\mathrm{A,theor} = dX_\mathrm{A,ipr} + (1-d)X_\mathrm{A,BKG},
$$
assuming that a fraction $d$ of the abundances $X_\mathrm{A,ipr}$ from an
i-process nucleosynthesis site had been mixed with a $(1-d)$ fraction of
some background (BKG) abundances $X_\mathrm{A,BKG}$ in the inter-stellar medium (ISM) or in the i-process source 
before that mixture was accreted by the star.
Therefore, the theoretical abundances depend on the assumed background and initial abundances $X_\mathrm{A,init}$.

\subsection{The pinning method}

The predicted abundances in Figure \ref{fig:fig3} are computed under the assumptions that
$X_\mathrm{A,init}\propto X_\mathrm{Fe,obs}$ and $X_\mathrm{A,BKG} = 0$, as for the Population-III ISM.
We also assume that $X_\mathrm{Fe,theor} = X_\mathrm{Fe,init} = X_\mathrm{Fe,obs}$, because
the strong depletion of the Fe abundance in Figure \ref{fig:fig2} is an artefact of our
one-zone nucleosynthesis simulation, while in a multi-zone He shell of a real stellar i-process site
the Fe seed nuclei are replenished by convective mixing. In this case, we have
$$
[\mathrm{A}/\mathrm{Fe}]_\mathrm{theor} = \log_{10}d + \log_{10}\frac{X_\mathrm{A,ipr}}{X_{\mathrm{A,init}}}.
$$
The $[\mathrm{A}/\mathrm{Fe}]_\mathrm{theor}$ distribution is then normalized (pinned) to an element B 
using a normalization factor $d$, such that $[\mathrm{B}/\mathrm{Fe}]_\mathrm{theor} = [\mathrm{B}/\mathrm{Fe}]_\mathrm{obs}$.

\subsection{The dilution method}

\begin{figure}
        \centering
        \includegraphics[width=\columnwidth]{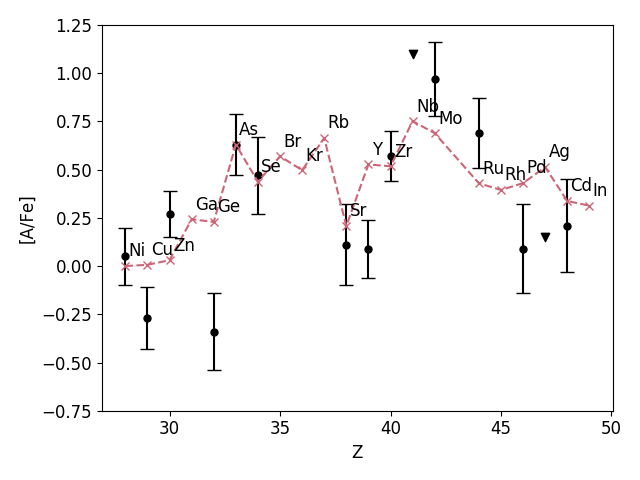}
        \caption{Same as the brown-dashed curve in Figure \ref{fig:fig3}, but using the dilution abundance comparison method.}
        \label{fig:dilute_method}
\end{figure}

If we assume that $X_\mathrm{A,init}$ is still proportional to $X_\mathrm{Fe,obs}$, but $X_\mathrm{A,BKG} = X_\mathrm{A,init}$ and
$X_\mathrm{Fe,theor}$ is now computed like the abundances of all other elements then
$$
[\mathrm{A}/\mathrm{Fe}]_\mathrm{theor} = \log_{10}\frac{\left[d\frac{X_\mathrm{A,ipr}}{X_{\mathrm{A},\odot}} + 
(1-d)\frac{X_\mathrm{A,init}}{X_{\mathrm{A},\odot}}\right]}{\left[d\frac{X_\mathrm{Fe,ipr}}{X_{\mathrm{Fe},\odot}} +
(1-d)\frac{X_\mathrm{Fe,init}}{X_{\mathrm{Fe},\odot}}\right]}.
$$
In this case, the factor $d$ plays the role of a dilution coefficient in mixing of the i-process abundances with the same initial
abundances that were used in the i-process nucleosynthesis simulations. This method is more suitable for a situation when
the i-process abundances were accreted locally from a binary or stellar cluster companion of the star.
The best-fit theoretical abundance distribution from Figure \ref{fig:fig3} is plotted in Figure \ref{fig:dilute_method} using the dilution
method. The important difference between the two comparison methods is that only the pinning method can
provide negative theoretical abundances that sometimes are actually observed, like the Cu and Ge abundances in HD94028.

\bsp	
\label{lastpage}
\end{document}